\documentclass[11pt,twoside,onecolumn]{article}
\usepackage[]{latexsym}
\usepackage{epsfig}
\usepackage{amsmath,amssymb}
\RequirePackage[dvipsnames,usenames]{color}
\newcommand{\be}{\begin{eqnarray}}
\newcommand{\ee}{\end{eqnarray}}

\title{\bf Anisotropic solutions by gravitational decoupling}
\author{J Ovalle$^{ab}$\thanks{Corresponding author: jovalle@usb.ve}
$\,$ 
R Casadio$^{cd}$\thanks{casadio@bo.infn.it}
$\,$
R da Rocha$^{e}$\thanks{roldao.rocha@ufabc.edu.br} 
$\,$
A Sotomayor$^{f}$\thanks{adrian.sotomayor@uantof.cl}
\\
\null
\\
$^a${\em Institute of Physics and Research Centre of Theoretical Physics and Astrophysics,}
\\
{\em Faculty of Philosophy and Science, Silesian University in Opava}
\\
{\em CZ-746 01 Opava, Czech Republic}
\\
$^b${\em Departamento de F\'{\i}sica, Universidad Sim\'on Bol\'ivar,}
\\
{\em AP 89000, Caracas 1080A, Venezuela}
\\
$^c${\em Dipartimento di Fisica e Astronomia, Alma Mater Universit\`a di Bologna}
\\
{\em via Irnerio~46, 40126 Bologna, Italy}
\\
$^d${\em Istituto Nazionale di Fisica Nucleare, Sezione di Bologna, I.S.~FLAG}
\\
{\em  viale Berti~Pichat~6/2, 40127 Bologna, Italy}
\\
$^e${\em Centro de Matem\'atica, Computa\c c\~ao e Cogni\c c\~ao,}
\\
{\em Universidade Federal do ABC (UFABC)}
\\
{\em  09210-580, Santo Andr\'e, SP, Brazil.}
\\
$^f${\em Departamento de Matem\'aticas, Universidad de Antofagasta}
\\
{\em  Antofagasta, Chile}
}
\begin{document}
\maketitle
\begin{abstract}
We investigate the extension of isotropic interior solutions for static self-gravitating systems to
include the effects of anisotropic spherically symmetric gravitational sources by means of the
gravitational decoupling realised via the minimal geometric deformation approach.
In particular, the matching conditions at the star surface with the outer Schwarzschild
space-time are studied in great detail, and we describe how to generate, from a single
physically acceptable isotropic solution, new families of anisotropic solutions whose
physical acceptability is also inherited from their isotropic parent.
\end{abstract}
%
%\keywords{Brane-world, Stars, Black holes}
%\pacs{04.50.+h, 04.70.-s, 04.70.Dy}
%
%
%
%
%
%
%\newpage
\section{Introduction}
\setcounter{equation}{0}
In a recent paper~\cite{MGD-decoupling}, the first simple, systematic and direct approach to
decoupling gravitational sources in general relativity (GR) was developed from the so-called
Minimal Geometric Deformation (MGD) approach.
The MGD was originally proposed~\cite{jo1,jo2} in the context of the Randall-Sundrum
brane-world~\cite{lisa1,lisa2} and extended to investigate
new black hole solutions~\cite{MGDextended1,MGDextended2}
(For some earlier works on MGD, see for instance Refs.~\cite{jo6,jo8,jo9,jo10},
and for some recent applications Refs.~\cite{jo11,jo12,roldaoGL,rrplb,rr-glueball, rr-acustic,Nicolini}).
The decoupling of gravitational sources by MGD (henceforth MGD-decoupling) is not only a novel idea,
but also has a number of ingredients that make it particularly attractive in the search for new
spherically symmetric solutions of Einstein's field equations, as discussed below.
\par
The main and foremost feature of this approach is that we can start from a simple
spherically symmetric gravitational source $\hat T_{\mu\nu}$ and add to it 
more and more complex gravitational sources, as long as the spherical symmetry is
preserved.
The starting source $\hat T_{\mu\nu}$ could be as simple as we wish, including the vacuum indeed,
to which we can add a first new source, say
\be
\label{coupling}
\hat T_{\mu\nu}\mapsto \tilde T^{(1)}_{\mu\nu}=\hat T_{\mu\nu}+\alpha^{(1)}\,T^{(1)}_{\mu\nu}
\ ,
\ee
and then repeat the process with more sources, namely
\be
\tilde T^{(1)}_{\mu\nu}\mapsto \tilde T^{(2)}_{\mu\nu}=\tilde T^{(1)}_{\mu\nu}+\alpha^{(2)}\,T^{(2)}_{\mu\nu}
\ ,
\ee
and so on.
In this way, we can extend straightforward solutions of the Einstein equations associated
with the simplest gravitational source $\hat T_{\mu\nu}$ into the domain of more intricate
forms of gravitational sources $T_{\mu\nu}=\tilde T^{(n)}_{\mu\nu}$, step by step and systematically. 
We stress that this method works as long as the sources do not exchange energy-momentum among them,
namely
\be
\nabla_{\nu}\hat T^{\mu\nu}
=
\nabla_{\nu} T^{(1)\mu\nu}
=
\ldots
=
\nabla_{\nu} T^{(n)\mu\nu}
=
0
\ ,
\ee
which further clarifies that the constituents shown in Eq.~(\ref{coupling}) can only couple via gravity.
\par
The converse also works.
In order to find a solution to Einstein's equations with a complex spherically symmetric energy-momentum
tensor $T_{\mu\nu}$, we can split it into simpler components, say $T^{(i)}_{\mu\nu}$,
and solve Einstein's equations for each one of these parts.
Hence, we will have as many solutions as are the contributions $T^{(i)}_{\mu\nu}$ in the original
energy-momentum tensor.
Finally, by a straightforward combination of all these solutions, we will obtain the solution to the
Einstein equations associated with the original energy-momentum tensor $T_{\mu\nu}$.
\par
To summarise, the MGD-decoupling amounts to the following procedure:
given two gravitational sources A and B, standard Einstein's equations are first solved for A,
and then a simpler set of quasi-Einstein equations are solved for B.
Finally, the two solutions can be combined in order to derive the complete solution for the total
system A$\cup$B.
Since Einstein's field equations are non-linear, the above procedure represents a breakthrough
in the search and analysis of solutions, especially when the involved situations are beyond trivial
cases, such as the interior of self-gravitating systems dominated by gravitational sources more
realistic than the ideal perfect fluid~\cite{lake2,visser2005}.
Of course, we remark that this decoupling occurs because of the spherical symmetry and
time-independence of the systems under investigation. 
\par
Although decoupling gravitational sources in GR by a systematic way represents
in itself a fact of significant theoretical importance, its practical relevance is no less.
Indeed, this simple and systematic method could be conveniently exploited in a large number
of relevant cases, such as the Einstein-Maxwell~\cite{EM} and Einstein-Klein-Gordon
system~\cite{shapiro,bosonstars,konstantin,singleton}, for higher derivative
gravity~\cite{stelle,sengupta,roman}, $f(R)$-theories of
gravity~\cite{diFelice,thomas, salvatore,salvatore2,marcelo1,marcelo2,Alvarez},
Ho$\check{\rm r}$ava-aether gravity~\cite{carloni2017,ted}, polytropic spheres~\cite{zdenek1,zdenek2,zdenek3}, among many others.
In this respect, the simplest practical application of the MGD-decoupling consists in
extending known isotropic and physically acceptable interior solutions for 
spherically symmetric self-gravitating systems into the anisotropic domain,
at the same time preserving physical acceptability, which represents a
highly non-trivial problem~\cite{Stephani}
(for obtaining anisotropic solutions in a generic way, see for instance
Ref.~\cite{lake3,luis1,sharma}).
\par
The paper is organised as follows:
in Section~\ref{s2}, we briefly review the effective Einstein field equations
for a spherically symmetric and static distribution of matter with effective density
$\tilde{\rho}$, effective radial pressure $\tilde{p}_r$ and effective tangential pressure
$\tilde{p}_t$;
Section~\ref{s3} is devoted to the fundamentals of the MGD-decoupling;
in Section~\ref{s4}, we provide detail on the matching conditions under the
MGD-decoupling;
in Section~\ref{s5}, we implement the MGD-decoupling to extend
perfect fluid solutions in the anisotropic domain.
In particular, three new exact and physically acceptable anisotropic solutions, 
generated from a single perfect fluid solution, are developed
in order to emphasise the robustness of the approach;
finally, we summarise our conclusions in Section~\ref{s6}.
\section{Einstein equations for multiple sources}
\label{s2}
\setcounter{equation}{0}
\par
Let us start from the standard Einstein field equations 
\begin{equation}
\label{corr2}
R_{\mu\nu}-\frac{1}{2}\,R\, g_{\mu\nu}
=
-k^2\,T^{\rm (tot)}_{\mu\nu}
\ ,
\end{equation}
with
\begin{equation}
\label{emt}
T^{\rm (tot)}_{\mu\nu}
=
T^{\rm (m)}_{\mu\nu}+\alpha\,\theta_{\mu\nu}
\ ,
\end{equation}
where
\begin{equation}
\label{perfect}
T^{\rm (m)}_{\mu \nu }=(\rho +p)\,u_{\mu }\,u_{\nu }-p\,g_{\mu \nu }
\end{equation}
is the 4-dimensional energy-momentum tensor of a perfect fluid with 4-velocity field $u^\mu$,
density $\rho$ and isotropic pressure $p$.
The term $\theta_{\mu\nu}$ in Eq.~(\ref{emt}) describes any additional source whose coupling to gravity
is further proportional to the constant $\alpha$~\cite{Matt}.
This source may contain new fields, like scalar, vector and tensor fields,
and will in general produce anisotropies in self-gravitating systems.
We just recall that, since the Einstein tensor is divergence free, the total energy-momentum tensor~(\ref{emt})
must satisfy the conservation equation
\begin{equation}
\nabla_\nu\,T^{{\rm (tot)}{\mu\nu}}=0
\ .
\label{dT0}
\end{equation}
\par 
In Schwarzschild-like coordinates, a static spherically symmetric metric reads 
\begin{equation}
ds^{2}
=
e^{\nu (r)}\,dt^{2}-e^{\lambda (r)}\,dr^{2}
-r^{2}\left( d\theta^{2}+\sin ^{2}\theta \,d\phi ^{2}\right)
\ ,
\label{metric}
\end{equation}
where $\nu =\nu (r)$ and $\lambda =\lambda (r)$ are functions of the areal
radius $r$ only, ranging from $r=0$ (the star center) to some $r=R$ (the
star surface), and the fluid 4-velocity is given by $u^{\mu }=e^{-\nu /2}\,\delta _{0}^{\mu }$
for $0\le r\le R$.
The metric~(\ref{metric}) must satisfy the Einstein equations~(\ref{corr2}),
which explicitly read
\begin{eqnarray}
\label{ec1}
-k^2
\left(
\rho+\alpha\,\theta_0^{\,0}
\right)
&\!\!=\!\!&
-\frac 1{r^2}
+
e^{-\lambda }\left( \frac1{r^2}-\frac{\lambda'}r\right)\ ,
\\
\label{ec2}
-k^2
\left(-p+\alpha\,\theta_1^{\,1}\right)
&\!\!=\!\!&
-\frac 1{r^2}
+
e^{-\lambda }\left( \frac 1{r^2}+\frac{\nu'}r\right)\ ,
\\
\label{ec3}
-k^2
\left(-p+\alpha\,\theta_2^{\,2}\right)
&\!\!=\!\!&
\frac {e^{-\lambda }}{4}
\left( 2\,\nu''+\nu'^2-\lambda'\,\nu'
+2\,\frac{\nu'-\lambda'}r\right)
\ .
\end{eqnarray}
The conservation equation~(\ref{dT0}), which is a linear combination of Eqs.~(\ref{ec1})-(\ref{ec3}), yields
\begin{equation}
\label{con1}
p'
+
\frac{\nu'}{2}\left(\rho+p\right)
-
\alpha\left(\theta_1^{\,\,1}\right)'
+
\frac{\nu'}{2}\alpha\left(\theta_0^{\,\,0}-\theta_1^{\,\,1}\right)
+
\frac{2\,\alpha}{r}\left(\theta_2^{\,\,2}-\theta_1^{\,\,1}\right)
=
0
\ ,
\end{equation}
where $f'\equiv \partial_r f$.
We then note the perfect fluid case is formally recovered for $\alpha\to 0$.
\par
The system~(\ref{ec1})-(\ref{ec3}) contains seven unknown functions, namely:
two physical variables, the density $\rho(r)$ and pressure $p(r)$;
two geometric functions, the temporal metric function $\nu(r)$
and the radial metric function $\lambda(r)$;
and three independent components of $\theta_{\mu\nu}$.
These equations therefore form an indefinite system.
In the particular case where $\theta_{\mu\nu}$ depends only on the density and the pressure,
we need to prescribe only an additional equation to close the system Eqs.~(\ref{ec1})-(\ref{ec3}),
just as we do for the perfect fluid in standard GR.
However, at this stage we want to emphasise that it is not enough to know the space-time geometry
to determine the gravitational source $\{\rho, p, \theta_{\mu\nu}\}$ in general.
\par
In order to simplify the analysis of the system~(\ref{ec1})-(\ref{ec3}), and by simple inspection, we
can identify an effective density 
\begin{equation}
\tilde{\rho}
=
\rho
+\alpha\,\theta_0^{\,0}
\ ,
\label{efecden}
\end{equation}
an effective isotropic pressure
\begin{equation}
\tilde{p}_{r}
=
p-\alpha\,\theta_1^{\,1}
\ ,
\label{efecprera}
\end{equation}
and an effective tangential pressure
\begin{equation}
\tilde{p}_{t}
=
p-\alpha\,\theta_2^{\,2}
\ .
\label{efecpretan}
\end{equation}
These definitions clearly illustrate that the source $\theta_{\mu\nu}$ could in general induce an
anisotropy,
\begin{equation}
\label{anisotropy}
\Pi
\equiv
\tilde{p}_{t}-\tilde{p}_{r}
=
\alpha\left(\theta_1^{\,1}-\theta_2^{\,2}\right)
\ ,
\end{equation}
inside the stellar distribution.
The system of Eqs.~(\ref{ec1})-(\ref{ec3}) could indeed be treated as an anisotropic fluid~\cite{Luis,tiberiu},
which would require to consider five unknown functions, namely, the two metric functions
$\nu(r)$ and $\lambda(r)$, and the effective functions in Eqs.~(\ref{efecden})-(\ref{efecpretan}). 
However, we are going to implement a different approach, as explained below. 
\section{Gravitational decoupling by MGD}
\setcounter{equation}{0}
\label{s3}
We shall implement the MGD in order to solve the system of Eqs.~(\ref{ec1})-(\ref{con1}).
In this approach, the system will be transformed in such a way that the field equations
associated with the source $\theta_{\mu\nu}$ will take the form of the
``effective quasi-Einstein'' Eqs.~(\ref{ec1d})-(\ref{ec3d}).
\par
Let us start by considering a solution to Eqs.~(\ref{ec1})-(\ref{con1}) with $\alpha=0$,
namely, a GR perfect fluid solution $\{\xi,\mu,\rho,p\}$, where $\xi$ and $\mu$
are the corresponding metric functions.
The metric~(\ref{metric}) now reads
\begin{equation}
ds^{2}
=
e^{\xi (r)}\,dt^{2}
-
\frac{dr^{2}}{\mu(r)}
-
r^{2}\left( d\theta^{2}+\sin ^{2}\theta \,d\phi ^{2}\right)
\ ,
\label{pfmetric}
\end{equation}
where 
\begin{equation}
\label{standardGR}
\mu(r)
\equiv
1-\frac{k^2}{r}\int_0^r x^2\,\rho\, dx
=1-\frac{2\,m(r)}{r}
\end{equation}
is the standard GR expression containing the mass function $m$.
Now let us turn on the parameter $\alpha$ to consider the effects of the
source $\theta_{\mu\nu}$ on the perfect fluid solution $\{\xi,\mu\,\rho,p\}$.
These effects can be encoded in the geometric deformation undergone by the
perfect fluid geometry $\{\xi,\mu\}$ in Eq.~(\ref{pfmetric}), namely
\begin{eqnarray}
\label{gd1}
\xi
&\mapsto &
\nu
=
\xi+\alpha\,g
\ ,
\\
\label{gd2}
\mu 
&\mapsto &
e^{-\lambda}
=
\mu+\alpha\,f
\ ,
\end{eqnarray}
where $g$ and $f$ are the deformations undergone by the temporal and radial
metric component, respectively.
Among all possible deformations~(\ref{gd1}) and (\ref{gd2}), the so-called minimal geometric
deformation is given by
\begin{eqnarray}
\label{gd11}
g
&\mapsto &
0
\\
\label{gd22}
f
&\mapsto &
f^*
\ .
\end{eqnarray}
In this case, the metric in Eq.~(\ref{pfmetric}) is minimally deformed by $\theta_{\mu\nu}$ and its
radial metric component becomes
\begin{eqnarray}
\label{expectg}
\mu(r)\mapsto\,e^{-\lambda(r)}
=
\mu(r)+\alpha\,f^{*}(r)
\ ,
\end{eqnarray}
whereas the temporal metric component $e^{\nu}$ remains unchanged.
(More precisely, $\nu(r)$ becomes $\nu(r,\alpha)$, as can be seen, for instance, in Eq.~(\ref{matchmimic})
below.)
Upon replacing Eq.~(\ref{expectg}) in the Einstein equations~(\ref{ec1})-(\ref{ec3}),
the system splits into two sets of equations:
\par
\noindent
i) the first one is given by the standard Einstein equations for a perfect fluid (the one with $\alpha = 0$ we
started from), whose metric is given by Eq.~(\ref{pfmetric}) with $\xi(r) = \nu(r)$:
\begin{eqnarray}
\label{ec1pf}
k^2\rho
&\!\!=\!\!&
\frac{1}{r^2}
-\frac{\mu}{r^2}
-\frac{\mu'}{r}\ ,
\\
\label{ec2pf}
k^2\,p
&\!\!=\!\!&
-\frac 1{r^2}+\mu\left( \frac 1{r^2}+\frac{\nu'}r\right)\ ,
\\
\label{ec3pf}
k^2\,p
&\!\!=\!\!&
\frac{\mu}{4}\left(2\nu''+\nu'^2+\frac{2\nu'}{r}\right)+\frac{\mu'}{4}\left(\nu'+\frac{2}{r}\right)
\ ,
\end{eqnarray}
along with the conservation equation~(\ref{dT0}) with $\alpha = 0$, namely $\nabla_\nu\,T^{{\rm (m)}{\mu\nu}}=0$,
yielding
\begin{equation}
\label{conpf}
p'+\frac{\nu'}{2}\left(\rho+p\right) = 0
\ ,
\end{equation}
which is a linear combination of Eqs~(\ref{ec1pf})-(\ref{ec3pf}).
And
\par
\noindent
ii) the second set contains the source $\theta_{\mu\nu}$ and reads
\begin{eqnarray}
\label{ec1d}
k^2\,\theta_0^{\,0}
&\!\!=\!\!&
-\frac{f^{*}}{r^2}
-\frac{f^{*'}}{r}
\ ,
\\
\label{ec2d}
k^2\,\theta_1^{\,1}
&\!\!=\!\!&
-f^{*}\left(\frac{1}{r^2}+\frac{\nu'}{r}\right)
\ ,
\\
\label{ec3d}
k^2\,\theta_2^{\,2}
&\!\!=\!\!&
-\frac{f^{*}}{4}\left(2\nu''+\nu'^2+2\frac{\nu'}{r}\right)
-\frac{f^{*'}}{4}\left(\nu'+\frac{2}{r}\right)
\ .
\end{eqnarray}
The conservation equation~(\ref{dT0}) then yields to $\nabla_\nu\,\theta^{\mu\nu}=0$, which explicitly reads
\begin{equation}
\label{con1d}
\left(\theta_1^{\,\,1}\right)'
-\frac{\nu'}{2}\left(\theta_0^{\,\,0}-\theta_1^{\,\,1}\right)
-\frac{2}{r}\left(\theta_2^{\,\,2}-\theta_1^{\,\,1}\right)
=
0
\ .
\end{equation}
Eqs.~\eqref{conpf} and~\eqref{con1d} simply mean that there is no exchange of energy-momentum between
the perfect fluid and the source $\theta_{\mu\nu}$, so that their interaction is purely gravitational. 
\par 
As was previously remarked in Ref.~\cite{MGD-decoupling}, Eqs.~(\ref{ec1d})-(\ref{con1d})
look very similar to the standard spherically symmetric Einstein field equations for an anisotropic fluid
with energy-momentum tensor $\theta_{\mu\nu}$, that is
$\{{\rho}= \theta_0^{\,0};\,\, {p}_{r}=-\theta_1^{\,1};\,\, {p}_{t}=-\theta_2^{\,2}\}$.
The corresponding metric would be given by
\begin{equation}
ds^{2}
=
e^{\nu (r)}\,dt^{2}-\frac{dr^{2}}{f^{*}(r)}
-r^{2}\left( d\theta^{2}+\sin ^{2}\theta \,d\phi ^{2}\right)
\ .
\label{metricaniso}
\end{equation}
However, the right-hand sides of Eqs.~(\ref{ec1d}) and~(\ref{ec2d}) are not the standard expressions
for the Einstein tensor components $G_0^{\,\,0}$ and $G_1^{\,\,1}$, since there is a missing $-1/r^2$
in both.
Nonetheless, the system of Eqs.~(\ref{ec1d})-(\ref{con1d}) may be formally identified as Einstein equations
for an anisotropic fluid with energy-momentum tensor ${\theta}^*_{\mu\nu}$, whose effective energy
density $\tilde{\rho}$, effective isotropic pressure $\tilde{p}_r$ and effective tangential pressure $\tilde{p}_t$
are given respectively by
\begin{eqnarray}
\label{theta1}
\tilde{\rho}
&\!\!=\!\!&
{\theta^*}_0^{\,\,0}=\theta_0^{\,0}+\frac{1}{k^2\,r^2}
\ ,
\\
\label{theta2}
\tilde{p}_r
&\!\!=\!\!&
{\theta^*}_1^{\,\,1}=\theta_1^{\,1}+\frac{1}{k^2\,r^2}
\ ,
\\
\label{theta3}
\tilde{p}_t
&\!\!=\!\!&
{\theta^*}_2^{\,\,2}=\theta_2^{\,2}={\theta^*}_3^{\,\,3}=\theta_3^{\,3}
\ ,
\end{eqnarray}
which can be written more concisely as
\begin{equation}
\label{shift2}
k^2\,\theta_\mu^{*\,\,\nu}
=
k^2\,\theta_\mu^{\,\nu}+\frac{1}{r^2}\left(\delta_\mu^{\,\,\,0}\,\delta_0^{\,\,\,\nu}+\delta_\mu^{\,\,\,1}\,\delta_1^{\,\,\,\nu}\right)
\ .
\end{equation}
With the transformation~(\ref{shift2}), the conservation equation~(\ref{con1d}) takes the standard form
\begin{equation}
\label{con1dd}
\left({\theta}_1^{*\,1}\right)'
-\frac{\nu'}{2}\left({\theta}_0^{*\,0}-{\theta}_1^{*\,1}\right)
-\frac{2}{r}\left({\theta}_2^{*\,2}-{\theta}_1^{*\,1}\right)
=
0
\end{equation}
and therefore the interpretation of Eqs.~(\ref{ec1d})-(\ref{con1d}) as standard spherically symmetric
Einstein equations for the source $\theta^{*}_{\mu\nu}$ in Eqs.~(\ref{theta1})-(\ref{theta3})
is now complete.
\par
As was pointed out in Ref.~\cite{MGD-decoupling}, since Eqs.~(\ref{ec1d}) and~(\ref{ec2d})
do not contain the standard Einstein tensor components, we should expect that the conservation
equation~(\ref{con1d}) for the source $\theta_{\mu\nu}$ is not longer a linear combination of
Eqs.~(\ref{ec1d})-(\ref{ec3d}).
However, Eq.~(\ref{con1d}) still remains a linear combination of the system~(\ref{ec1d})-(\ref{ec3d}).
The MGD approach therefore turns the indefinite system~(\ref{ec1})-(\ref{ec3}) into the set of equations
for a perfect fluid $\{\rho,p,\nu,\mu\}$ plus a simpler system of four unknown functions
$\{f^{*},\,\theta_0^{\,0},\,\theta_1^{\,1},\,\theta_2^{\,2}\}$ satisfying the three
Eqs.~(\ref{ec1d})-(\ref{ec3d}) [at this stage we suppose that we have already found a perfect fluid solution,
thus $\nu$ is determined], or the equivalent anisotropic system of Eqs.~(\ref{theta1})-(\ref{theta3}).
Either way, the system~(\ref{ec1})-(\ref{ec3}) has been successfully decoupled.
\par 
The MGD-decoupling can be summarised in a more formal way as follows: 
consider a static spherically symmetric gravitational source ${T}_{\mu\nu}$ containing one isotropic
perfect fluid $\hat T_{\mu\nu}$ and $n$ other gravitational sources $T_{\mu\nu}^{(i)}$, namely
\begin{equation}  
\label{MGD1}
T_{\mu\nu}
=
\hat T_{\mu\nu}
+
\sum_{i=1}^{n}\,\alpha_i\,T_{\mu\nu}^{(i)}
\ ,
\end{equation}
then the diagonal metric $g_{\mu\nu}$, solution of Einstein equation $G_{\mu\nu}=-k^2\,T_{\mu\nu}$, reads
\begin{eqnarray}
g_{\mu\nu}
&\!\!=\!\!&
\hat{g}_{\mu\nu}={g}_{\mu\nu}^{(i)}
\qquad
{\rm for}
\quad
\mu=\nu\neq\,1
\ ,
\\
g^{11}
&\!\!=\!\!&
\hat{g}^{11}+\alpha_1\,g^{(1)11}+\cdots+\alpha_n\,g^{(n)11}
\ .
\end{eqnarray}
This metric $g_{\mu\nu}$ is found by first solving the Einstein equations for the source $\hat{T}_{\mu\nu}$ ,
\begin{equation}
\hat{G}_{\mu\nu}=-k^2\,\hat{T}_{\mu\nu}\ ,
\qquad\qquad
\nabla_\nu\,\hat{T}^{{\mu\nu}}=0\ ,
\end{equation}
and then by solving the remaining $n$ quasi-Einstein's equations for the sources $T_{\mu\nu}^{(i)}$,
namely
\begin{eqnarray}
\tilde{G}_{\mu\nu}^{(1)}
&\!\!=\!\!&
-k^2\,{T}_{\mu\nu}^{(1)}
\ ,
\qquad\qquad
\nabla_\nu\,{T}^{{(1)\mu\nu}}=0
\ ,
\nonumber
\\
&\vdots &
\nonumber
\\
\tilde{G}_{\mu\nu}^{(n)}
&\!\!=\!\!&
-k^2\,{T}_{\mu\nu}^{(n)}\ ,
\qquad\qquad
\nabla_\nu\,{T}^{{(n)\mu\nu}}=0
\ ,
\label{MGD2}
\end{eqnarray}
where the divergence-free quasi-Einstein tensor $\tilde{G}_{\mu\nu}$ is related with the standard ${G}_{\mu\nu}$
by 
\begin{equation}
\tilde{G}_{\mu}^{\,\,\,\,\nu}
=
{G}_{\mu}^{\,\,\,\,\nu}+\Gamma_{\mu}^{\,\,\,\,\nu}(g)
\ ,
\end{equation}
with $\Gamma_{\mu}^{\,\,\,\,\nu}(g)$ a tensor that depends exclusively on ${g}_{\mu\nu}$ to ensure
the divergence-free condition.
In the spherically symmetric representation it reads
\begin{equation}
\label{Gamma}
\Gamma_{\mu}^{\,\,\,\,\nu}
=
\frac{1}{r^2}\left(\delta_\mu^{\,\,\,0}\,\delta_0^{\,\,\,\nu}+\delta_\mu^{\,\,\,1}\,\delta_1^{\,\,\,\nu}\right)
\ .
\end{equation}
The explicit components of $\tilde{G}_{\mu}^{\,\,\,\,\nu}$ in terms of the metric in Eq.~(\ref{metricaniso})
are shown in the right-hand side of Eqs.~(\ref{ec1d})-(\ref{ec3d}). 
\section{Matching condition for stellar distributions}
\label{s4}
\setcounter{equation}{0}
A crucial aspect in the study of stellar distributions is the matching conditions
at the star surface ($r=R$) between the interior ($r<R$)  and the exterior ($r>R$)
space-time geometries~\cite{israel}.
In our case, the interior stellar geometry is given by the MGD metric 
\begin{equation}
ds^{2}
=
e^{\nu^{-}(r)}\,dt^{2}
-\left(1-\frac{2\,\tilde{m}(r)}{r}\right)^{-1}dr^2
-r^{2}\left(d\theta ^{2}+\sin {}^{2}\theta d\phi ^{2}\right)
\ ,
\label{mgdmetric}
\end{equation}
where the interior mass function is given by
\begin{equation}
\label{effecmass}
\tilde{m}(r)
=
m(r)-\frac{r}{2}\,\alpha\,f^{*}(r)
\ , 
\end{equation}
with $m$ given by the standard GR expression in Eq.~(\ref{standardGR}) and
$f^{*}$ the yet to be determined MGD in Eq.~(\ref{expectg}).
\par
The inner metric~(\ref{mgdmetric}) should now be matched with an outer geometry 
where there is no isotropic fluid, that is $p^+=\rho^+=0$.
The exterior $r>R$ may however not be a vacuum anymore since we can, in general,
have new fields coming from the sector described by $\theta_{\mu\nu}$.
The general outer metric can be written as
\begin{equation}
ds^{2}
=
e^{\nu^{+}(r)}\,dt^{2}
-e^{\lambda^{+}(r)}\,dr^{2}
-r^{2}\left(d\theta ^{2}+\sin {}^{2}\theta d\phi ^{2}\right)
\ ,
\label{genericext}
\end{equation}
where the explicit form of the functions $\nu ^{+}$ and $\lambda ^{+}$ are
obtained by solving the effective 4-dimensional exterior Einstein equations
\begin{equation}
\label{exterior}
R_{\mu \nu }-\frac{1}{2}\,R^\alpha_{\ \alpha}\,g_{\mu \nu}
= \alpha\,
\theta_{\mu \nu }
\ .
\end{equation} 
The MGD will reduce the exterior Einstein equations~\eqref{exterior} to the
system~(\ref{ec1d})-(\ref{ec3d}), where the geometric function $\nu$ is given by the
Schwarzschild solution instead of a perfect fluid solution.
\par
Continuity of the first fundamental form at the star surface $\Sigma$ 
defined by $r=R$ reads
\begin{equation}
\left[ ds^{2}\right] _{\Sigma }=0
\ ,
\label{match1}
\end{equation}
where 
$[F]_{\Sigma }\equiv F(r\rightarrow R^{+})-F(r\rightarrow R^{-})\equiv F_{R}^{+}-F_{R}^{-}$,
for any function $F=F(r)$, which yields 
\begin{equation}
{\nu ^{-}(R)}
=
{\nu ^{+}(R)}
\ ,
\label{ffgeneric1}
\end{equation}
and
\begin{equation}
1-\frac{2\,M_0}{R}+\alpha\,f^{*}_{R}
=
e^{-\lambda ^{+}(R)}
\ ,
\label{ffgeneric2}
\end{equation}
where $M_0=m(R)$ and $f^{*}_{R}$ is the minimal geometric deformation at the star surface.
\par
Likewise, continuity of the second fundamental form reads
\begin{equation}
\left[G_{\mu \nu }\,r^{\nu }\right]_{\Sigma }
=
0
\ ,
\label{matching1}
\end{equation}
where $r_{\mu }$ is a unit radial vector.
Using Eq.~(\ref{matching1}) and the general Einstein equations~(\ref{corr2}),
we then find 
\begin{equation}
\left[T_{\mu \nu }^{\rm (tot)}\,r^{\nu }\right]_{\Sigma}
=
0
\ ,
\label{matching2}
\end{equation}
which leads to 
\begin{equation}
\left[ p-\alpha\,\theta_1^{\,\,1}\right]_{\Sigma }
=
0
\ .
\label{matching3}
\end{equation}
This matching condition takes the final form 
\begin{equation}
p_{R}
-\alpha\,(\theta_1^{\,\,1})^{-}_{R}
=
-\alpha\,(\theta_1^{\,\,1})^{+}_{R}
\ ,
\label{matchingf}
\end{equation}
where $p_{R}\equiv p^{-}(R)$. 
The condition in Eq.~(\ref{matchingf}) is the general expression for the second fundamental form associated
with the Einstein equations given by Eq.~(\ref{corr2}).
\par
By using Eq.~(\ref{ec2d}) for the interior geometry in the condition~(\ref{matchingf}), the second fundamental
form can be written as
\begin{equation}
p_{R}
+\alpha\,\frac{f_{R}^{*}}{k^2}\left(\frac{1}{R^{2}}+\frac{\nu _{R}^{\prime }}{R}\right)
=
-\alpha\,(\theta_1^{\,\,1})^{+}_{R}
\ ,
\label{sfgeneric}
\end{equation}
where $\nu _{R}^{\prime }\equiv \partial _{r}\nu^{-}|_{r=R}$. 
Moreover, using now Eq.~(\ref{ec2d}) for the outer geometry in Eq.~(\ref{sfgeneric}) yields
\begin{equation}
p_{R}
+\alpha\,\frac{f_{R}^{*}}{k^2}\left(\frac{1}{R^{2}}+\frac{\nu _{R}^{\prime }}{R}\right)
=
\alpha\,\frac{g_{R}^{*}}{k^2}\left[\frac{1}{R^{2}}+\frac{2\,{\cal M}}{R^3}\frac{1}{(1-\frac{2\,{\cal M}}{R})}\right]
\ ,
\label{sfgenericf}
\end{equation}
where $g_{R}^{*}$ is the geometric deformation for the outer Schwarzschild solution due to the source
$\theta_{\mu\nu}$, that is
\begin{equation}
\label{Schw}
ds^2=\left(1-\frac{2{\cal M}}{r}\right)dt^2-\left(1-\frac{2{\cal M}}{r}+\alpha\,g^{*}(r)\right)^{-1}dr^2-d\Omega^2
\ .
\end{equation}
Eqs.~(\ref{ffgeneric1}), (\ref{ffgeneric2}) and~(\ref{sfgenericf}) are the necessary
and sufficient conditions for the matching of the interior MGD metric~(\ref{mgdmetric})
to a spherically symmetric outer ``vacuum'' described by the deformed Schwarzschild
metric in Eq.~({\ref{Schw}}).
Note that this exterior could be filled by fields contained in the source $\theta_{\mu\nu}$.
\par
The matching condition~(\ref{sfgenericf}) yields an important result:
if the outer geometry is given by the exact Schwarzschild metric, one must have $g^{*}(r) = 0$ in Eq.~(\ref{Schw}),
which then leads to the condition
\begin{equation}
\tilde{p}_R\equiv\,p_{R}+\alpha\,\frac{f_{R}^{\ast }}{k^2}
\left(\frac{1}{R^{2}}+\frac{\nu _{R}^{\prime }}{R}\right)=0
\ .
\label{pnegative}
\end{equation}
It is important to remark that the star will therefore be in equilibrium in a true (Schwarzschild) vacuum only if the
effective (in general anisotropic radial) pressure at the surface vanishes.
In particular, if the inner geometric deformation $f^*(r<R)$ is positive and weakens the gravitational field
[see Eq.~(\ref{effecmass})], an outer Schwarzschild vacuum can only be compatible with a non-vanishing
inner $\theta_{\mu\nu}$ if the isotropic stellar matter has $p_{R}<0$ at the star surface. 
This may be interpreted as regular matter with a solid crust~\cite{jo11}.
Otherwise, the condition $p_{R}=0$ can be obtained by imposing that the right-hand side of
Eq.~(\ref{pnegative}) be proportional to $p_R$, namely, $\alpha\,(\theta_1^{\,\,1})^{-}_{R}\sim\,p_{R}$
in Eq.~(\ref{matchingf}), which leads to a vanishing inner deformation $f^*_R=0$. 
\section{Interior solutions}
\label{s5}
\setcounter{equation}{0}
Let us now solve the Einstein's field Eqs.~(\ref{ec1})-(\ref{ec3}) for the interior of a self-gravitating anisotropic
system by the MGD decoupling.
The physical variables $\{\tilde{\rho}, \tilde{p}_r, \tilde{p}_t\}$ defined by Eqs.~(\ref{efecden})-(\ref{efecpretan})
and the two metric functions $\{\nu,\lambda\}$ in Eq.~(\ref{metric}) will be derived.
The first step is to turn off $\alpha$ and find a solution for the perfect fluid Einstein Eqs.~(\ref{ec1pf})-(\ref{ec3pf}).
In particular, we can simply choose a known solution with physical relevance, like the well-known Tolman~IV solution
$\{\nu,\mu,\rho, p\}$ for perfect fluids~\cite{Tolman}, namely 
\begin{eqnarray}
\label{tolman00}
e^{\nu(r)}
&\!\!=\!\!&
B^2\,\left(1+\frac{r^2}{A^2}\right)
\ ,
\\
\label{tolman11}
\mu(r)
&\!\!=\!\!&
\frac{\left(1-\frac{r^2}{C^2}\right)\left(1+\frac{r^2}{A^2}\right)}{1+\frac{2\,r^2}{A^2}}
\ ,
\\
\label{tolmandensity}
\rho(r)
&\!\!=\!\!&
\frac{3A^4+A^2\left(3C^2+7r^2\right)+2 r^2 \left(C^2+3 r^2\right)}{k^2\,C^2\left(A^2+2r^2\right)^2}
\ ,
\end{eqnarray}
and
\begin{equation}
\label{tolmanpressure}
p(r)
=
\frac{C^2-A^2-3r^2}{k^2\,C^2\left(A^2+2r^2\right)}
\ .
\end{equation}
The constants $A$, $B$ and $C$ in Eqs.~(\ref{tolman00})-(\ref{tolmanpressure}) are determined from the
matching conditions in Eqs.~(\ref{match1}) and (\ref{matching1}) between the above interior solution and
the exterior metric which we choose to be the Schwarzschild space-time.
This yields
\begin{equation}
\label{A}
\frac{A^2}{R^2}
=
\frac{R-3\,M_0}{M_0}
\ ,
\qquad
B^2=1-\frac{3\,M_0}{R}
\ ,
\qquad
\frac{C^2}{R^2}
=
\frac{R}{M_0}
\ ,
\end{equation}
with the compactness $M_0/R<4/9$, and $M_0=m(R)$ the total mass in Eq.~(\ref{standardGR}).
The expressions in Eq.~(\ref{A}) ensure the geometric continuity at $r=R$ and will change when we
add the source $\theta_{\mu\nu}$.
\par
Let us then turn $\alpha$ on in the interior.
The temporal and radial metric components are given by Eqs.~(\ref{tolman00}) and~(\ref{expectg}) respectively,
where the interior deformation $f^*(r)$ and the source $\theta_{\mu\nu}$ are related through Eqs.~(\ref{ec1d})-(\ref{ec3d}).
Hence, we need to prescribe additional information in order to close the system of Eqs.~(\ref{ec1d})-(\ref{ec3d}).
We have many alternatives for this purpose, like imposing an equation of state for the source $\theta_{\mu\nu}$
or some physically motivated restriction on $f^*(r)$.
In any case, we must be careful in keeping the physical acceptability of our solution, which is not a trivial issue.
In the following, this problem is addressed in such a way that three new, exact and physically acceptable
interior solutions will be generated.
\begin{figure}[t]
\center
\includegraphics[scale=0.3]{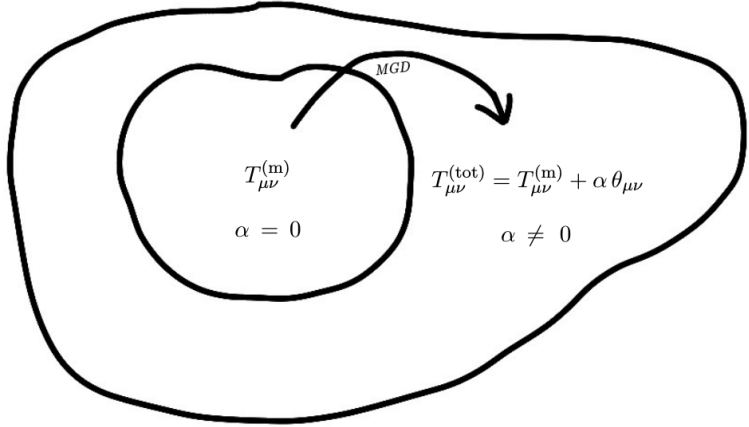}
\\
\caption{Any perfect fluid solution can be consistently extended to the anisotropic domain via the MGD-decoupling.}
\label{ext}      
\end{figure}
% 
%
%
%
%
%%%%%%%%%%%%%%%
\subsection{Solution~I: mimic constraint for pressure}
\label{s5.1}
From the matching condition (\ref{pnegative}) we see that the Schwarzschild exterior solution will be
compatible with regular interior matter as long as $\alpha\,(\theta_1^{\,\,1})^{-}_{R}\sim p_{R}$.
The simplest choice satisfying this critical requirement is
\begin{equation}
\label{constraint}
\theta_1^{\,\,1}(r)
=
p(r)
\ ,
\end{equation}
which, according to Eq.~(\ref{ec2pf}), can be written as
\begin{equation}
\label{mimic2}
k^2\,\theta_1^{\,1}
=
-\frac{1}{r^2}+\mu(r)\left(\frac{1}{r^2}+\frac{\nu'}{r}\right)
\ .
\end{equation}
From Eq.~(\ref{ec2d}) we can see that the ``mimic'' constraint in Eq.~(\ref{constraint}) is equivalent to
\begin{equation}
\label{mimic}
f^*(r)
=
-\mu(r)+\frac{1}{1+r\,\nu'(r)}
\ .
\end{equation}
Hence the radial metric component reads
\begin{equation}
\label{tolman11d}
e^{-\lambda(r)}
=
\left(1-\alpha\right)\mu(r)
+\alpha\left(\frac{A^2+r^2}{A^2+3\,r^2}\right)
\ ,
\end{equation}
where the expressions in Eqs.~(\ref{expectg}) and (\ref{tolman00}) have been used.
The interior metric functions given by Eqs.~(\ref{tolman00}) and (\ref{tolman11d})
represent the Tolman~IV solution minimally deformed by the generic anisotropic source
$\theta_{\mu\nu}$.
We can see that the limit $\alpha\to 0$ in Eq.~(\ref{tolman11d}) leads to the standard Tolman~IV
solution for perfect fluids. This is represented by a generic way in Fig.~\ref{ext}.
\par
Now let us match our interior metric in Eq.~(\ref{metric}) with metric functions~(\ref{tolman00})
and (\ref{tolman11d}) with the exterior Schwarzschild solution~({\ref{Schw}}) with $g^{*}(r)=0$.
We can see that, for a given distribution of mass $M_0$ and radius $R$, we have four unknown parameters,
namely $\{A, B, C\}$ from the interior solution in Eqs.~(\ref{tolman00}) and (\ref{tolman11d}), and the mass
${\cal M}$ in Eq.~({\ref{Schw}}).
Since we have only the three matching conditions~(\ref{ffgeneric1}), (\ref{ffgeneric2})
and~(\ref{matchingf}) at the star surface, the problem is not closed.
An obvious solution would be to set $B=1$ in Eq.~(\ref{tolman00}), corresponding to the time rescaling 
$t\to\tilde{t}=B\,t$.
However, we want to keep $B$ near its expression in the Tolman~IV solution of
Eqs.~(\ref{tolman00})-(\ref{tolmanpressure}) in order to see clearly the effect of the anisotropic source
$\theta_{\mu\nu}$ on the perfect fluid.
We will therefore solve for $\{A, C, {\cal M}\}$ with respect to $B$ as shown further below.
\par
The continuity of the first fundamental form given by Eqs.~(\ref{ffgeneric1}) and (\ref{ffgeneric2})
leads to
\begin{equation}
\label{fff1}
B^2\left(1+\frac{R^2}{A^2}\right)
=
1-\frac{2{\cal M}}{R}
\ ,
\end{equation}
and
\begin{equation}
\label{fff2}
\left(1-\alpha\right)\mu(R)
+\alpha\left(\frac{A^2+R^2}{A^2+3\,R^2}\right)
=
1-\frac{2{\cal M}}{R}
\ ,
\end{equation}
whereas continuity of the second fundamental form in Eq.~(\ref{matchingf}) yields
\begin{equation}
\label{sff}
p_{R}
-\alpha\,(\theta_1^{\,\,1})^{-}_{R}
=0\ .
\end{equation}
By using the mimic constraint in Eq.~(\ref{constraint}) in the condition (\ref{sff}), we obtain 
\begin{equation}
\label{sfff}
p_R=0
\ ,
\end{equation}
which, according to the expression in Eq.~(\ref{tolmanpressure}), leads to
\begin{equation}
\label{C}
C^2=A^2+3\,R^2
\ .
\end{equation}
On the other hand, by using the condition in Eq.~(\ref{fff2}), we obtain for the Schwarzschild mass
\begin{equation}
\label{MASS}
\frac{2{\cal M}}{R}
=
\frac{2\,M_0}{R}+\alpha\left(1-\frac{2\,M_0}{R}\right)-\alpha\left(\frac{A^2+R^2}{A^2+3\,R^2}\right)
\ ,
\end{equation}
where the expression in Eq.~(\ref{standardGR}) has been used.
Finally, by using the expression in Eq.~(\ref{MASS}) in the matching condition~(\ref{fff1}), we obtain
\begin{equation}
\label{matchmimic}
B^2\left(1+\frac{R^2}{A^2}\right)
=
(1-\alpha)\left(1-\frac{2\,M_0}{R}\right)
+\alpha\left(\frac{A^2+R^2}{A^2+3\,R^2}\right)
\ .
\end{equation}
\par
\begin{figure}[t]
\center
\includegraphics[scale=0.70]{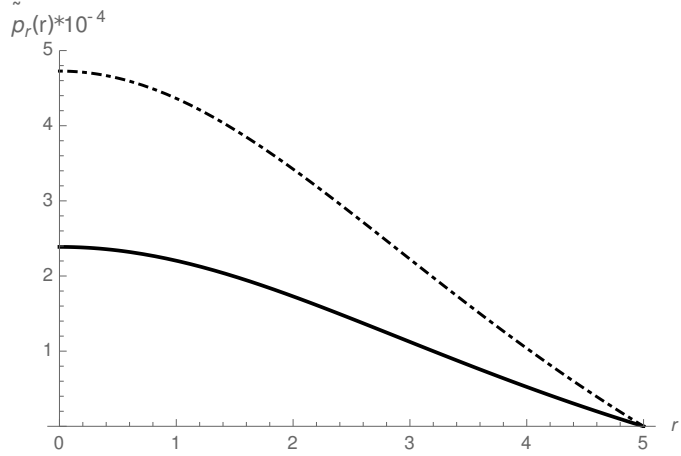}
\\
\caption{Solution~I:
effective radial pressure $\tilde{p}_r(r)\,[10^{-4}]$ for $B^2=2/5$ with $\alpha=0.5$ (lower curve) and $\alpha=0.01$ (upper curve)
for a stellar system with compactness $M_0/R=0.2$.}
\label{pre}      
\end{figure}
Eqs.~(\ref{C})-(\ref{matchmimic}) are the necessary and sufficient conditions to match the interior solution 
with the exterior Schwarzschild space-time.
By using the mimic constraint in Eq.~(\ref{constraint}) along with Eq.~(\ref{C}), the effective isotropic pressure
$\tilde{p}_r$ in Eq.~(\ref{efecprera}) reads
\begin{equation}
\label{pressrf}
\tilde{p}_r(r,\alpha)
=
\frac{3\,(1-\alpha)(R^2-r^2)}{k^2\,(A^2+3\,R^2)(A^2+2\,r^2)}
\ ,
\end{equation}
from which we see that the effective isotropic pressure mimics the (physically acceptable) perfect fluid pressure
$p(r)$ in Eq.~(\ref{tolmanpressure}).
On the other hand, the effective density and effective tangential pressure are given respectively by
\begin{eqnarray}
\label{denf}
\tilde{\rho}(r,\alpha)
&\!\!=\!\!&
(1-\alpha)\,\rho(r)
+\frac{6\,\alpha\,(A^2+r^2)}{k^2\,(A^2+3\,r^2)^2}
\ ,
\\
\label{pretanf}
\tilde{p}_t(r,\alpha)
&\!\!=\!\!&
\tilde{p}_r(r,\alpha)
+\frac{3\,{\alpha}\,r^2}{k^2\,(A^2+3\,r^2)^2}
\ .
\end{eqnarray}
Eqs.~(\ref{tolman00}) and (\ref{tolman11d}) along with Eqs.~(\ref{pressrf})-(\ref{pretanf}) represent
an exact Tolman~IV analytic solution to the system of Eqs.~(\ref{ec1})-(\ref{ec3}) minimally deformed
by the anisotropic source $\theta_{\mu\nu}$.
In particular, according to Eq.~(\ref{anisotropy}), the source $\theta_{\mu\nu}$ generates an anisotropy
given by
\begin{equation}
\Pi(r,\alpha)
=
\frac{3\,\alpha\,r^2}{k^2(A^2+r^2)^2}
\ .
\end{equation}
We can further solve Eq.~\eqref{matchmimic} for $A=A(B,\alpha)$, with $B$ as a free parameter
around the value in Eq.~(\ref{A}), in order to obtain the effective radial pressure $\tilde{p}_r(r,\alpha)$ in
Eq.~(\ref{pressrf}) shown in Figure~\ref{pre} for two values of $\alpha$ and $B^2=2/5$.
It appears that the anisotropy produced by $\theta_{\mu\nu}$ decreases the effective radial pressure
more and more for increasing $\alpha$.
\subsection{Solution~II: mimic constraint for density}
\label{s5.2}
An alternative choice leading to physically acceptable solutions is the ``mimic constraint" for the density,
\begin{equation}
\label{Dmimic}
\theta_0^{\,\,0}
=
\rho
\ ,
\end{equation}
which yields the first order differential equation
\begin{equation}
f'(r)+\frac{f(r)}{r}
=
-r\,k^2\,\rho
\ .
\end{equation}
The solution is given by
\begin{equation}
\label{Df}
f(r)
=
\frac{C_0}{r}
-\frac{r^2\left(A^2+C^2+r^2\right)}{C^2\,\left(A^2+2\,r^2\right)}
\ ,
\end{equation}
where the density $\rho$ in Eq.~(\ref{tolmandensity}) has been used.
To avoid a singular behaviour at the center $r=0$ we must impose $C_0=0$, 
so that Eq.~(\ref{expectg}) yields
\begin{equation}
\label{D11}
e^{-\lambda(r)}
=
\mu(r)
-\alpha\,\frac{r^2\left(A^2+C^2+r^2\right)}{C^2\,\left(A^2+2\,r^2\right)}
\ .
\end{equation}
By using the expression in Eq.~(\ref{D11}), the matching conditions in (\ref{ffgeneric1}) and (\ref{ffgeneric2})
lead to 
\begin{equation}
\label{DMASS}
\frac{2{\cal M}}{R}
=
\frac{2\,M_0}{R}
+\alpha\,\frac{R^2}{C^2}\left(\frac{A^2+C^2+R^2}{A^2+2R^2}\right)
\end{equation}
and
\begin{equation}
\label{Dmatchmimic}
B^2\left(1+\frac{R^2}{A^2}\right)
=
1-\frac{2\,M_0}{R}-\alpha\,\frac{R^2}{C^2}\left(\frac{A^2+C^2+R^2}{A^2+2R^2}\right)
\ ,
\end{equation}
whereas continuity of the second fundamental form in Eq.~(\ref{sff}) yields
\begin{equation}
\label{DC}
C^2
=
\frac{(1+\alpha)(A^2+R^2)(A^2+3R^2)}{(A^2+R^2)-\alpha(A^2+3R^2)}
\ .
\end{equation}
\par
Eqs.~(\ref{DMASS})-(\ref{DC}) are the necessary and sufficient conditions to match the exterior Schwarzschild solution
with the interior solution in Eq.~(\ref{metric}) with metric functions given in Eqs.~(\ref{tolman00}) and (\ref{D11}).
By using these metric functions in the field equation~(\ref{ec2}), the effective isotropic pressure in Eq.~(\ref{efecprera})
reads
\begin{equation}
\label{DPr}
\tilde{p}_r(r,\alpha)
=
p(r)
-\alpha\,\frac{(A^2+C^2+r^2)(A^2+3\,r^2)}
{k^2\,C^2(A^2 + r^2) (A^2 + 2 r^2)}
\ .
\end{equation}
On the other hand, the mimic constraint for the density in Eq.~(\ref{Dmimic}) yields
\begin{equation}
\label{Dtolmandensity}
\tilde{\rho}(r,\alpha)
=
(1+\alpha)\,\rho(r)
\ ,
\end{equation}
whereas the effective tangential pressure reads
\begin{equation}
\label{Dpretanf}
\tilde{p}_t(r,\alpha)
=
\tilde{p}_r(r,\alpha)
+\frac{\alpha\,r^2}{k^2\,(A^2 + r^2)^2}
\ ,
\end{equation}
with the anisotropy thus given by
\begin{equation}
\label{DPI}
\Pi(r,\alpha)
=
\frac{\alpha\,r^2}{k^2\,(A^2 + r^2)^2}
\ .
\end{equation}
\par
\begin{figure}[t]
\center
\includegraphics[scale=0.65]{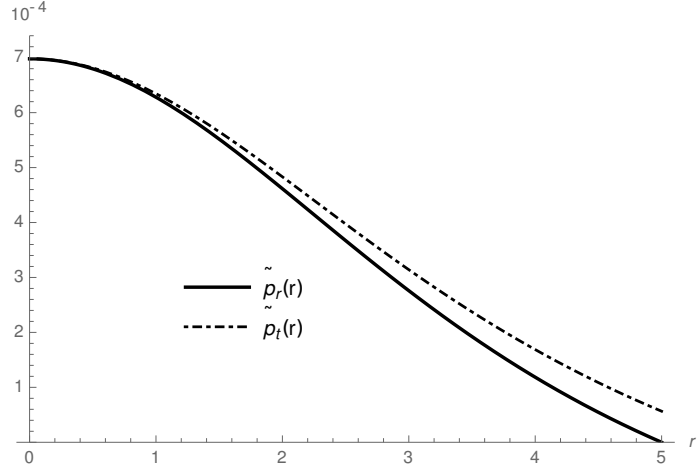}
\\
\caption{Solution~II: effective radial pressure $\tilde{p}_r(r,\alpha)$ and tangential pressure $\tilde{p}_t(r,\alpha)$
for a stellar system with compactness $M_0/R=0.2$ and $\alpha=0.2$.
We can see that $\Pi(r,\alpha)$ increases towards the surface.
This effect is proportional to $\alpha$, in agreement with Eq.~(\ref{DPI}).}
\label{Dpre}      
\end{figure}

We then employ the matching conditions~(\ref{Dmatchmimic}) and (\ref{DC}) to find $A=A(B,\alpha)$, 
again leaving $B$ as a free parameter with values around the expression in Eq.~(\ref{A}).
Figure~\ref{Dpre} shows the radial pressure $\tilde{p}_r(r,\alpha)$ and tangential pressure $\tilde{p}_t(r,\alpha)$
inside the stellar distribution, showing how the anisotropy $\Pi(r,\alpha)$ in Eq.~(\ref{DPI}) increases towards the surface.
\par
\begin{figure}[t]
\center
\includegraphics[scale=0.75]{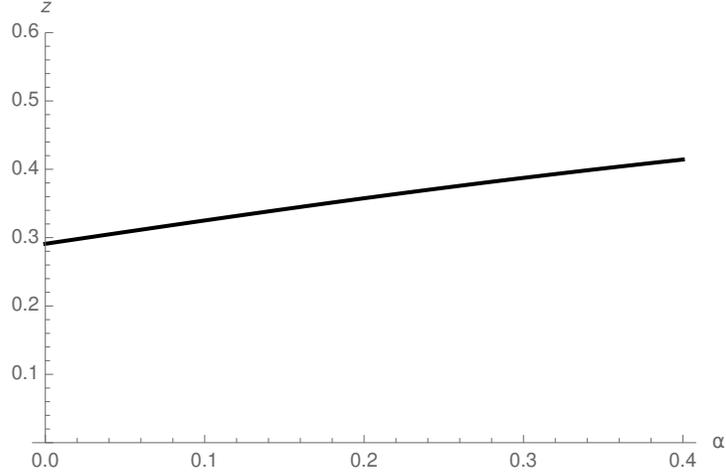}
\\
\caption{Solution~II: Gravitational redshift at the stellar surface $r=R$. We can see that the anisotropic effects increases the compactness of the self-gravitating system and therefore the surface redshift $z$.}
\label{redshift}      
\end{figure}

It is also interesting to investigate the effects of the anisotropy on the surface redshift~\cite{Ivanov},
\begin{equation}
\label{z}
z=\left[\frac{g_{00}(R)}{g_{00}(\infty)}\right]^{-1/2}-1
\ ,
\end{equation} 
which in our case reads
\begin{equation}
\label{z2}
z(\alpha)=\left[1-\frac{2\,{\cal M(\alpha)}}{R}\right]^{-1/2}-1
\ ,
\end{equation} 
and is displayed in Figure~\ref{redshift}.
In particular, we can see from the plot that the anisotropy increases the gravitational redshift at the stellar surface,
hence a distant observer will detect a more compact distribution in comparison with the isotropic case.
This result is in agreement with the fact that ${\cal M}> M_0$, as can be seen from Eq.~(\ref{DMASS}).
\subsection{Solution~III: extending anisotropic solutions}
So far we have seen how to generate exact and physically acceptable anisotropic solutions starting
from a known isotropic solution.
In order to emphasise the full potential of the MGD-decoupling, schematically represented by
Eqs.~(\ref{MGD1})-(\ref{Gamma}), we will use the anisotropic solution
$\{\nu,\lambda,\tilde{\rho},\tilde{p}_r,\tilde{p}_t\}$ from Section~\ref{s5.2}, and explicitly given by
Eqs.~(\ref{tolman00}), (\ref{D11}), (\ref{DPr})-(\ref{Dpretanf}), to generate a third anisotropic and physically
acceptable solution $\{\nu,\lambda,\bar{\rho},\bar{p}_r,\bar{p}_t\}$.
\par
Let us start by adding to the energy-momentum tensor in Eq.~(\ref{emt}) one more anisotropic source
with energy-momentum tensor $\psi_{\mu\nu}$, that is
\begin{equation}
T_{\mu\nu}\rightarrow\,\bar{T}_{\mu\nu}=\,T_{\mu\nu}+\beta\,\psi_{\mu\nu}\ ,
\end{equation}
where $\beta$ is a new coupling constant.
The Einstein equations now read
\begin{eqnarray}
\label{ec1A}
k^2
\left( \tilde{\rho} +\beta\,\psi_0^{\,0} \right)
&\!\!=\!\!&
\frac 1{r^2}
-e^{-\lambda }\left( \frac1{r^2}-\frac{\lambda'}r\right)
\ ,
\\
\label{ec2A}
k^2
\left(\tilde{p}_r-\beta\,\psi_1^{\,1}\right)
&\!\!=\!\!&
-\frac 1{r^2}+e^{-\lambda }\left( \frac 1{r^2}+\frac{\nu'}r\right)
\ ,
\\
\label{ec3A}
k^2
\left(\tilde{p}_t-\beta\,\psi_2^{\,2}\right)
&\!\!=\!\!&
\frac{ e^{-\lambda }}{4}\left[ 2\,\nu''+\nu'^2-\lambda'\,\nu'
+2\,\frac{\nu'-\lambda'}r\right]
\ ,
\end{eqnarray}
where $\{\tilde{\rho},\tilde{p}_r,\tilde{p}_t\}$ are shown in Eqs.~(\ref{DPr})-(\ref{Dpretanf}).
The temporal metric component $\nu(r)$ remains equal to the one in Eq.~(\ref{tolman00})
for the original Tolman~IV solution, whereas the radial metric component is deformed according to
\begin{equation}
\label{D11A}
e^{-\lambda(r)}
=
\mu(r)
-\alpha\,\frac{r^2\left(A^2+C^2+r^2\right)}{C^2\,\left(A^2+2\,r^2\right)}
+\beta\,g^{*}(r)
\ ,
\end{equation}
where $g^{*}(r)$ represents the geometric deformation.
After decoupling, the source $\psi_{\mu\nu}$ and the unknown deformation $g^{*}(r)$ satisfy
the same Eqs.~(\ref{ec1d})-(\ref{ec3d}), which now read
\begin{eqnarray}
\label{ec1dA}
k^2\,\psi_0^{\,0}
&\!\!=\!\!&
-\frac{g^{*}}{r^2}
-\frac{(g^{*})'}{r}\ ,
\\
\label{ec2dA}
k^2\,\psi_1^{\,1}
&\!\!=\!\!&
-g^{*}\left(\frac{1}{r^2}+\frac{\nu'}{r}\right)\ ,
\\
\label{ec3dA}
k^2\,\psi_2^{\,2}
&\!\!=\!\!&
-\frac{g^{*}}{4}\left(2\nu''+\nu'^2+2\frac{\nu'}{r}\right)
-\frac{(g^{*})'}{4}\left(\nu'+\frac{2}{r}\right)
\ .
\end{eqnarray}
We then impose the mimic constraint for pressure as
\begin{equation}
\psi_1^{\,1}=\tilde{p}_r
\ ,
\end{equation}
and following the same procedure as in Section~\ref{s5.1}, we obtain
\begin{equation}
\label{tolman00DA}
e^{-\lambda}
=
(1-\beta)\left[
\mu(r)
-\alpha\,\frac{r^2\left(A^2+C^2+r^2\right)}{C^2\,\left(A^2+2\,r^2\right)}\right]
+\beta\,\frac{A^2+r^2}{A^2+3\,r^2}
\ .
\end{equation}
\par
\begin{figure}[t]
\center
\includegraphics[scale=0.30]{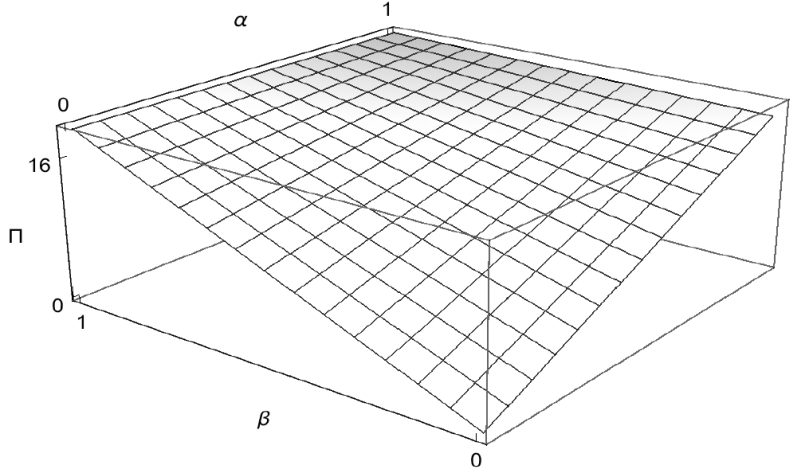}
\\
\centering\caption{Solution III: anisotropy $\Pi(r,\alpha,\beta)$ $x\,10^{5}$ at the stellar surface $r=R$.}
\label{sol3}      
\end{figure}

Continuity of the first fundamental form in Eq.~(\ref{match1}) again leads to the expression
in Eq.~(\ref{fff1}), where the mass ${\cal M}$ is now given by
\begin{equation}
\label{MASSA}
\frac{2{\cal M}}{R}
=
\frac{2\,M_0}{R}
+\beta\left(1-\frac{2\,M_0}{R}\right)
+(1-\beta)\,\alpha\,\frac{R^2\,\left(A^2+C^2+R^2\right)}{C^2\,\left(A^2+2\,R^2\right)}
-\beta\,\frac{A^2+R^2}{A^2+3\,R^2}
\ .
\end{equation}
Continuity of the second fundamental form in Eq.~(\ref{matching2}) again leads to (\ref{DC}). 
The new anisotropic solution to the Einstein Eqs.~(\ref{ec1dA})-(\ref{ec3dA}) is given by the Tolman~IV
temporal metric componen~(\ref{tolman00}), the radial metric component~(\ref{tolman00DA}), and the
three physical variables
\begin{eqnarray}
\label{barden}
\bar{\rho}(r,\alpha,\beta)
&\!\!=\!\!&
(1-\beta)(1+\alpha)\,\rho(r)
+\frac{6\,\beta\,(A^2+r^2)}{k^2\,(A^2+3\,r^2)^2}
\ ,
\\
\label{barPr}
\bar{p}_r(r,\alpha,\beta)
&\!\!=\!\!&
(1-\beta)
\left[
p(r)
-\alpha\,
\frac{(A^2+C^2+r^2)(A^2+3\,r^2)}{k^2\,C^2(A^2 + r^2) (A^2 + 2 r^2)}
\right]
\ ,
\\
\label{barPt}
\bar{p}_t(r,\alpha,\beta)
&\!\!=\!\!&
\bar{p}_r(r,\alpha,\beta)+\Pi(r,\alpha,\beta)
\ ,
\end{eqnarray}
with the anisotropy (see Fig.~\ref{sol3})
\begin{equation}
\label{barPI}
\Pi(r,\alpha,\beta)
=
\frac{r^2\left[3\beta\,(A^2+r^2)^2+\alpha\,(1-\beta)(A^2+3\,r^2)^2\right]}{k^2\,(A^2+r^2)^2(A^2+3\,r^2)^2}
\ .
\end{equation}
One can easily check that the solution~II of Section~\ref{s5.2} is the particular case $\beta=0$ of the solution~III
given by Eqs.~(\ref{tolman00}), (\ref{tolman00DA}) and (\ref{barden})-(\ref{barPt}).
Likewise, the solution~I of Section~\ref{s5.1} (with $\alpha\to\beta$) is the particular case $\alpha=0$ of the solution~III.
Eventually, we could set $\alpha=\beta$ in the solution~III so that the anisotropic sector is determined by a single free
parameter $\alpha$.
\section{Conclusions}
\label{s6}
By using the MGD-decoupling approach, we presented in detail how to extend interior isotropic solutions
for self-gravitating systems in order to include anisotropic (but still spherically symmetric) gravitational sources.
For this purpose, we showed that the Einstein field equations for a static and spherically symmetric self-gravitating
system in Eq.~(\ref{ec1})-(\ref{ec3}) can be decoupled in two sectors, namely:
the isotropic sector corresponding to a perfect fluid $\hat T_{\mu\nu}$ shown in Eq.~(\ref{ec1pf})-(\ref{ec3pf}), 
and the sector described by quasi-Einstein field equations associated with an anisotropic source
$\theta_{\mu\nu}$ shown in Eqs.~(\ref{ec1d})-(\ref{ec3d}).
These two sectors must interact only gravitationally, without direct exchange of energy-momentum.
\par
The matching conditions at the stellar surface were then studied in detail for an outer Schwarz\-schild space-time.
In particular, the continuity of the second fundamental form in Eq.~(\ref{pnegative}) was shown to yield the important result
that the effective radial pressure $\tilde{p}_R=0$.
The effective pressure~(\ref{efecprera}) contains both the isotropic pressure of the undeformed matter source
$\hat T_{\mu\nu}$ and the inner geometric deformation $f^*(r)$ induced by the energy-momentum $\theta_{\mu\nu}$.
We recall the latter could also represent a specific matter source, like a Klein-Gordon scalar field
or any other form of matter-energy, but also the induced effects of extra-dimensions in the brane-world.
If the geometric deformation $f^*(r)$ is positive and therefore weakens the gravitational field [see Eq.~(\ref{effecmass})],
an outer Schwarzschild vacuum can only be supported if the isotropic $p_{R}<0$ at the star surface.
This can in fact be interpreted as regular matter with a solid crust~\cite{jo11} as long as the region
with negative pressure does not extend too deep into the star.
\par
In order to show the robustness of our approach, three new exact and physically acceptable interior anisotropic
solutions to the Einstein field equations were generated from a single perfect fluid solution.
All these new solutions inherit their physical acceptability from the original isotropic solution.
In particular, it was shown that the anisotropic source $\theta_{\mu\nu}$ always reduces the isotropic radial
pressure $\tilde{p}(r)$ inside the self-gravitating system.
\par
We would like to remark that the MGD-decoupling is not just a technique for developing physically acceptable anisotropic
solutions in GR, but represents a powerful and efficient way to exploit the gravitational decoupling in relevant physical
problems.
The extension of GR solutions into the domain of more complex gravitational sources is a highly non-trivial theoretical
problem.
For instance, it is well known that a Klein-Gordon scalar field induces anisotropic effects when it is coupled with
the gravitational field through the Einstein equations. 
Hence the MGD-decoupling represents a useful tool for extending GR isotropic solutions for self-gravitating systems
into solutions of the Einstein-Klein-Gordon system.
It could be implemented, for instance, to investigate the role played by a Klein-Gordon scalar field
during the gravitational collapse.
In this respect, it is worth mentioning that the MGD-decoupling can be generalised for time-dependent scenarios,
as long as the spherical symmetry is preserved under slowly evolving situations, which means the stellar system
is always in hydrostatic equilibrium~\cite{Luis}.
\subsubsection*{Acknowledgements}
\par
J.O.~is supported by Institute of Physics and Research Centre of Theoretical Physics and Astrophysics,
Silesian University in Opava.
R.C.~is partially supported by the INFN grant FLAG.
R.dR.~is grateful to CNPq (Grant No. 303293/2015-2),
and to FAPESP (Grant No.~2017/18897-8), for partial financial support. 
A.S.~is partially supported by Project Fondecyt 1161192, Chile.

\end{document}